\documentclass[12pt]{article}

\renewcommand{\baselinestretch}{1.3}

\newcommand {\dimass} {$m(pK^0)$}
\newcommand {\cosdec} {$\cos\Theta_K^\mathrm{cms}$}
\newcommand {\abscosdec} {$|\cos\Theta_K^\mathrm{cms}|$}
\newcommand {\alotra} {$p_T < 300$ MeV and $p_L > 0$}
\newcommand {\window} {$445 < p_\mathrm{beam} < 535$ MeV}
\newcommand {\peak} {$1530 < m(pK^0) < 1546$ MeV}
\newcommand {\wings} {$1514 < m(pK^0) < 1530$ MeV and $1546 < m(pK^0) < 1562$ MeV}
\newcommand {\one} {$-0.6 < \cos\Theta_K^\mathrm{cms} < +0.2$}
\newcommand {\two} {$-0.7 < \cos\Theta_K^\mathrm{cms} < +0.3$}
\newcommand {\three} {$-0.75 < \cos\Theta_K^\mathrm{cms} < +0.35$}
\newcommand {\askew} {$dW/d\cos\Theta_K^\mathrm{cms} \sim (\cos\Theta_K^\mathrm{cms}+a)^2 + b$}
\newcommand {\symme} {$dW/d\cos\Theta_K^\mathrm{cms} \sim \cos^2\Theta_K^\mathrm{cms}$}
\begin{document}

\title
{ Angular distribution in $s$-channel formation of the pentaquark $\Theta^+$ baryon }

\author {
DIANA Collaboration\\
V.V. Barmin$^{1,\ast}$,    
A.E. Asratyan$^{1,\ast}$, 
C. Curceanu$^2$, 
G.V. Davidenko$^1$, \\
C. Guaraldo$^2$,
M.A. Kubantsev$^{1,\ast}$, 
I.F. Larin$^1$,
V.A. Matveev$^1$, \\
V.A. Shebanov$^1$,
N.N. Shishov$^1$, 
L.I. Sokolov$^1$,  
and V.V. Tarasov$^1$ \\
{\normalsize $^1$ \it Institute of Theoretical and Experimental Physics,
Moscow 117218, Russia}\\
{\normalsize $^2$ \it Laboratori Nazionali di Frascati dell' INFN,
C.P. 13-I-00044 Frascati, Italy} \\
{\normalsize $^\ast$ E-mail: barmin@itep.ru, ashot.asratyan@gmail.com, 
Mikhail.Kubantsev@gmail.com}
}              
\maketitle

\begin{abstract}
     Using the DIANA data on the charge-exchange reaction 
$K^+n \rightarrow pK^0$ on a bound neutron, in which the $s$-channel 
formation of the pentaquark baryon $\Theta^+(1538)$ has been observed,
we analyze the dependence of the background-subtracted 
$\Theta^+ \rightarrow pK^0$ signal on the $K^0$ emission angle in the 
$pK^0$ rest frame. In order to describe the observed 
$\cos\Theta_K^\mathrm{cms}$ distribution, invoking the interference 
between the nonresonant $s$-wave and the $\Theta^+$-mediated higher-wave 
contributions to the amplitude of the charge-exchange reaction is 
required at a 2.8$\sigma$ level. The spin--parity assignment of 1/2$^-$ 
for the $\Theta^+$ baryon is ruled out at a statistical level of 2.9 
standard deviations. A physically-meaningful selection 
in $\cos\Theta_K^\mathrm{cms}$ based on the observed
angular dependence of the $\Theta^+ \rightarrow pK^0$ signal allows to
boost the statistical significance of the signal up to 7.1 standard
deviations. This is far in excess of previously reported signals
and renders the $\Theta^+$ existence more credible.
\end{abstract}
PACS number(s): 13.75.Jz, 25.80.Nv

\newpage

     The exotic baryons with minimum quark configuration of
$(4q)\bar{q}$ have been theoretically discussed ever since the
emergence of the quark model \cite{Gell-Mann, Jaffe}.
For such objects 
formed of light quarks, the lowest $SU(3)$ representation was identified 
as the anti-decuplet that involves a single state with positive 
strangeness --- the isosinglet baryon $\Theta^+(uudd\bar{s})$. This 
pentaquark baryon can be uniquely identified by the $KN$ decays 
($K^+n$ and $K^0p$) that are forbidden for the three-quark baryons.
Since the ``fall-apart" mechanism is not suppressed by any obvious
selection rules, the decay width of the $\Theta^+$ baryon was 
phenomenologically assumed to be rather big ($\Gamma \sim 100$ MeV).
The first rigorous predictions for the anti-decuplet of light
pentaquark baryons were formulated in the landmark analysis
\cite{DPP} based on the chiral quark-soliton model. According to 
these theoretical predictions, the anti-decuplet baryons have 
spin--parity of 1/2$^+$, and the mass of the isosinglet $\Theta^+$ 
baryon should be close to 1530 MeV. The predicted decay width of the 
$\Theta^+$ proved to be far below the earlier phenomenological 
estimates: $\Gamma < 15$ MeV. Subsequently, some theorists using 
different assumptions came to a conclusion that the $\Theta^+$ 
decay width should be well below this upper limit --- on the order 
of 1 MeV or even less \cite{Lorce, Oganesian, Ledwig}.

     Narrow peaks near 1540 MeV in the effective-mass spectra of
the systems  $nK^+$ and $pK^0$ were initially observed in the reaction
$\gamma n \rightarrow n K^+ K^-$  on the $^{12}$C nucleus in the LEPS 
experiment \cite{Nakano-old}, and in the charge-exchange reaction
$K^+ n \rightarrow pK^0$ on the Xe nucleus in the DIANA experiment
\cite{DIANA-2003}. Subsequently, both experiments confirmed their 
initial observations \cite{Nakano, DIANA-2007, DIANA-2010, DIANA-2014}.
Using the dynamics of $s$-channel formation of the $\Theta^+$ in the
charge-exchange reaction $K^+ n \rightarrow pK^0$, DIANA was able to
directly probe the $\Theta^+$ decay width: $\Gamma = 0.34\pm0.10$ MeV
assuming $J = 1/2$.
Other searches for the $\Theta^+$ baryon in different reactions and
experimental conditions yielded both positive and negative results,
see the review papers \cite{Burkert,Danilov-Mizuk,Hicks}.
A number of experimental groups have reneged on their initial positive
evidence, that anyway was statistically insignificant and may have
resulted from wishful thinking and the so-called ``bandwagon effect".
Of the many null results, only a few that have been formulated in terms 
of the $\Theta^+$ intrinsic width should be treated as physically
meaningful. The best (albeit model-dependent) null result has been
reported by the E19 experiment at J-PARC, where the $\Theta^+$ signal 
was searched for in the $K^-$ missing mass in the hadronic reaction 
$\pi^- p \rightarrow K^- X$ \cite{E19}. The E19 upper limit on the 
$\Theta^+$ decay width, $\Gamma < 0.36$ MeV assuming the $\Theta^+$ 
spin--parity of 1/2$^+$, is narrowly consistent with the DIANA 
measurement. On the other hand, a group from the CLAS collaboration
has recently re-analyzed their data for the reaction
$\gamma p \rightarrow K^0_S K^0_L p$ on hydrogen, invoking
the interference between $\phi p$ and $\Theta^+ \bar{K^0}$ in the 
final state $p K^0_L K^0_S$ \cite{Amaryan}. A narrow 
statistically-significant peak near 1540 MeV, tentatively 
interpreted as the $\Theta^+$ signal, has been observed in the
$K^0_S$ missing-mass spectrum. This observation does not contradict
the null result earlier reported by CLAS for the same data sample
\cite{CLAS-null}. 

     In this paper, we continue the investigation of 
$\Theta^+$ formation in the charge-exchange reaction 
$K^+n \rightarrow pK^0$ on a bound neutron using the data of the 
DIANA experiment. In particular, we probe the angular distribution
of decay products in the $\Theta^+$ rest frame.
\begin{figure}[h]
\renewcommand{\baselinestretch}{1.}
\vspace { 7 cm }
\includegraphics{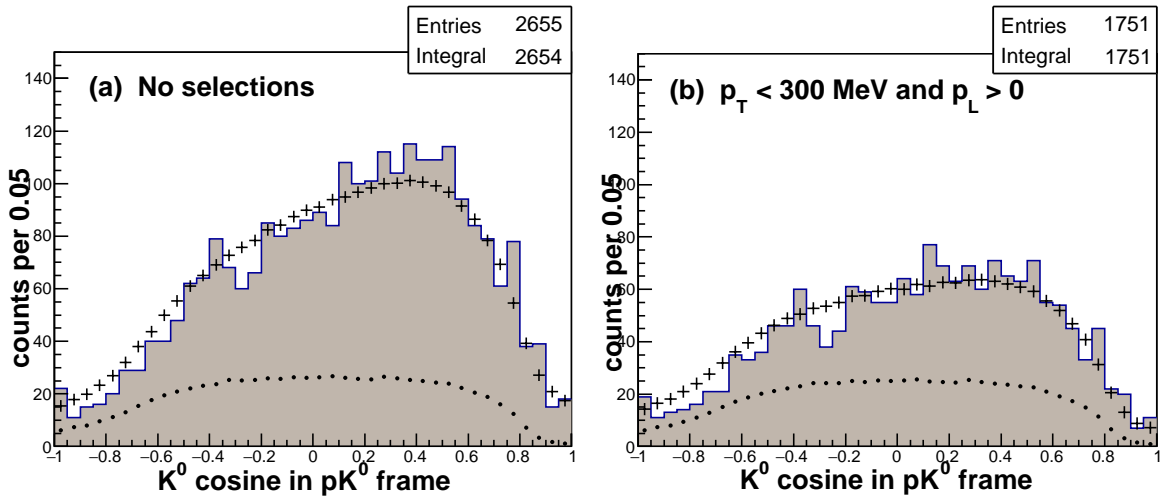}
\caption{
The cosine of the $K^0$ emission angle in the $pK^0$ rest frame, 
$\Theta_K^\mathrm{cms}$, for all measured events (a) and upon applying 
the selections \alotra\ (b). The corresponding distributions of 
simulated events are shown by crosses, and the contributions of 
rescattering-free events --- by dots.}
\label{cosdec}
\end{figure}

     The DIANA bubble chamber filled with liquid Xenon 
was exposed to a separated beam of monochromatic $K^+$ mesons from the
10-GeV proton synchrotron at ITEP, Moscow. In the fiducial volume of the 
bubble chamber, $K^+$ momentum varies from $\sim 730$ MeV for entering 
kaons to zero for those that range out through ionization. Throughout 
this momentum interval, all collisions and decays of incident $K^+$ 
mesons are efficiently detected. The $K^+$ momentum at interaction point 
is determined from the spatial distance between the detected vertex and 
the mean position of the vertices due to decays of stopping $K^+$ mesons. 
Charged secondaries (electrons, pions, kaons, and protons) are 
identified by ionization and by decays at rest for kaons, and 
momentum-analyzed by their range in Xenon. The detection efficiency 
for $\gamma$-quanta with $p_\gamma > 25$ MeV is close to 100\%. 
Secondary $K^0$ mesons are identified by the detectable decays 
$K^0_S \rightarrow \pi^+\pi^-$ and $K^0_S \rightarrow \pi^0\pi^0$, 
and momentum-analyzed using the kinematic reconstruction. (In this
analysis, only the former decay is used.) Further details on the 
experimental procedure may be found in \cite{DIANA-2014} and 
references therein.
The candidate events for the charge-exchange reaction
$K^+ n \rightarrow K^0 p$ with no intranuclear rescatterings are
selected as final states with a single proton and a
$K^0_S \rightarrow \pi^+\pi^-$ decay. The instrumental thresholds 
for the momenta of secondary particles are $p_K > 155$ MeV and 
$p_p > 165$ MeV. The experimental resolution is near 3.5 MeV for the 
$pK^0$ effective mass. 

     Plotted in Fig.~\ref{cosdec}(a) for all
measured events is the cosine of the $K^0$ emission angle in the 
$pK^0$ rest frame with respect to the $pK^0$ direction of motion,
\cosdec. Also shown is the \cosdec\ distribution of all simulated
$pK^0$ events (crosses) and of those in which the proton and the $K^0$
suffered no intranuclear rescatterings (dots). The former has
been normalized to the number of all measured $pK^0$ events.
The effect of the selections in the transverse and longitudinal 
momenta of the $pK^0$ system, \alotra, is shown in Fig.~\ref{cosdec}(b).
These are seen to reject the rescattered events rather than the 
unrescattered ones. The simulation procedure has been described in 
\cite{DIANA-2014}. 
\begin{figure}[!b]
\renewcommand{\baselinestretch}{1.}
\vspace { 11.5 cm }
\includegraphics{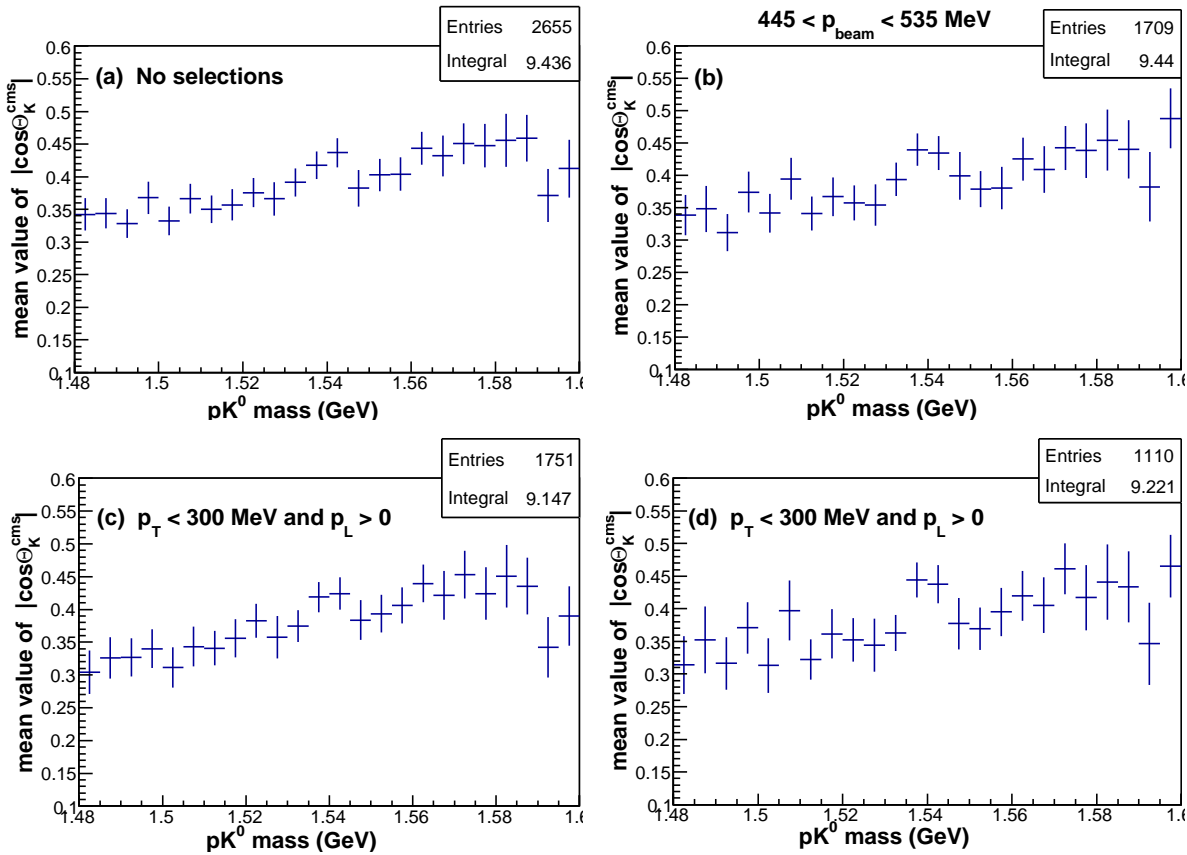}
\caption{
The mean value of \abscosdec\ as a function of the $pK^0$ effective 
mass for all measured events (a) and for those in the region \window\ (b). 
The effect of the selections \alotra\ is shown in (c) and (d).}
\label{profile}
\end{figure}

     The mean value of \abscosdec\ is plotted in Fig.~\ref{profile} as
a function of the $pK^0$ effective mass. The enhancement observed at
$m(pK^0) \simeq 1540$ MeV is emphasized by the selection in the 
$K^+$ momentum at interaction point, \window, that reflects the
dynamics of $s$-channel formation of the $\Theta^+$ baryon in the
reaction $K^+n \rightarrow pK^0$ on a bound neutron \cite{DIANA-2014}.
It is further emphasized by the selections \alotra\ aimed at rejecting 
the rescattered events. That the anomaly in \abscosdec\ occurs in the 
mass region of the observed $\Theta^+$ peak \cite{DIANA-2014} suggests 
that it is rooted in an ``anomalous" angular distribution of $\Theta^+$ 
decays that may show a quadratic term in \cosdec.
Therefore, it is interesting to compare the \cosdec\ distribution 
for the mass region of the peak with that for the sideband areas of 
\dimass. Except for the scatter plots discussed in the last paragraph, 
the selection \window\ is implicitly assumed throughout.
\begin{figure}[h]
\renewcommand{\baselinestretch}{1.}
\vspace { 5.5 cm }
\includegraphics{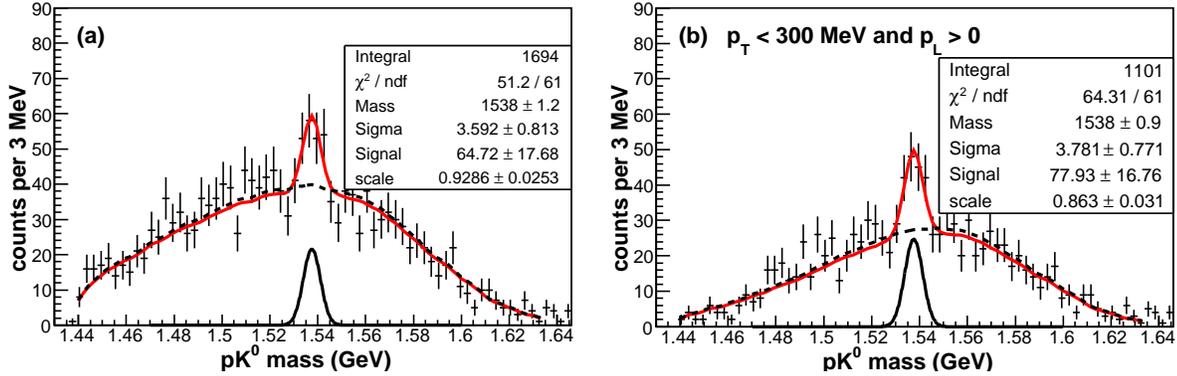}
\caption {
The $pK^0$ effective mass prior to (a) and upon (b) applying the
selections \alotra. Either mass spectrum is fitted to the 
simulated nonresonant background with variable normalization plus a Gaussian 
with variable position, width, and magnitude. The null fits to the background
form alone are shown by dashed lines.}
\label{dimass}
\end{figure}

\begin{figure}
\renewcommand{\baselinestretch}{1.}
\vspace { 11 cm }
\includegraphics{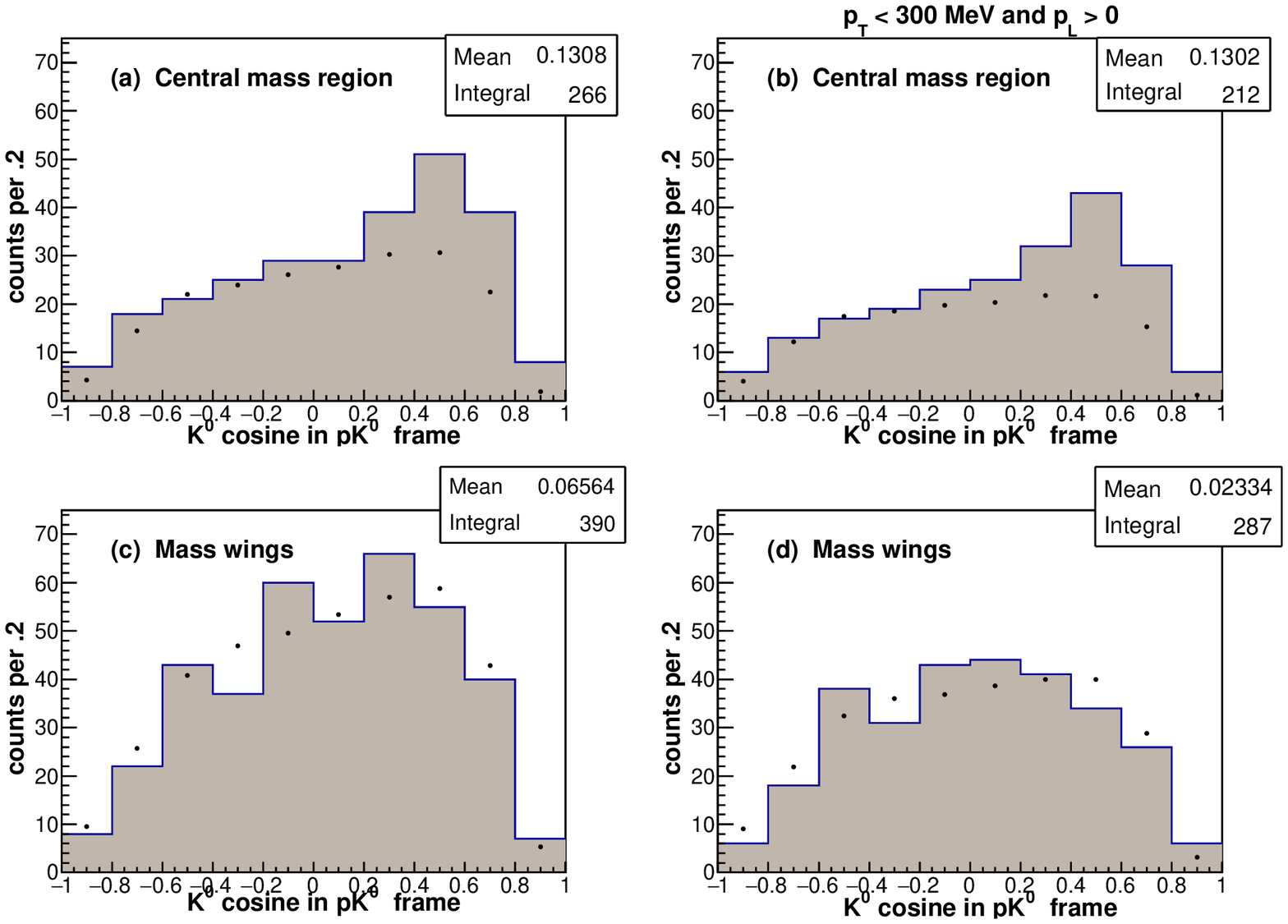}
\caption {
The \cosdec\ distributions for the $\Theta^+$ mass region of
\peak\ (a) and for the sideband regions of \wings\ (c). The effects of the 
selections \alotra\ are shown in (b) and (d). The simulated distributions 
are depicted by dots.}
\label{peak-wings}
\end{figure}

     The distribution of the $pK^0$ effective mass is shown in 
Fig.~\ref{dimass}(a), and upon applying the selections \alotra\ ---
in Fig.~\ref{dimass}(b). Either mass spectrum is then 
fitted to the simulated nonresonant background with variable normalization 
plus a Gaussian with variable position, width, and magnitude. The width
of the observed $\Theta^+$ peak near 1538 MeV is consistent with the
experimental resolution of $\sigma_m \simeq 3.5$ MeV. In agreement with 
the fitted width of the $\Theta^+$ signal, the peak area of the $pK^0$ 
effective mass is selected as \peak, and the sideband areas --- as \wings.
The \cosdec\ distributions of events in the peak and sideband areas are
shown in Figs. \ref{peak-wings}(a) and \ref{peak-wings}(c), respectively. The
effect of the selections \alotra\ is shown in the corresponding right-hand
panels. For the sideband areas, the simulated distribution (dots) is 
normalized to the observed one by the number of events. Then, the same 
scaling factor is applied to the simulated \cosdec\ distribution for 
the peak area. (As a result, there the simulated \cosdec\ spectrum runs
lower than the observed one.) The simulation that assumes a pure $s$-wave
for the nonresonant reaction $K^+n \rightarrow pK^0$ \cite{Dover-Walker}
agrees with the data for the sideband areas, but not for the peak area. 

\begin{figure}
\renewcommand{\baselinestretch}{1.}
\vspace { 6 cm }
\includegraphics{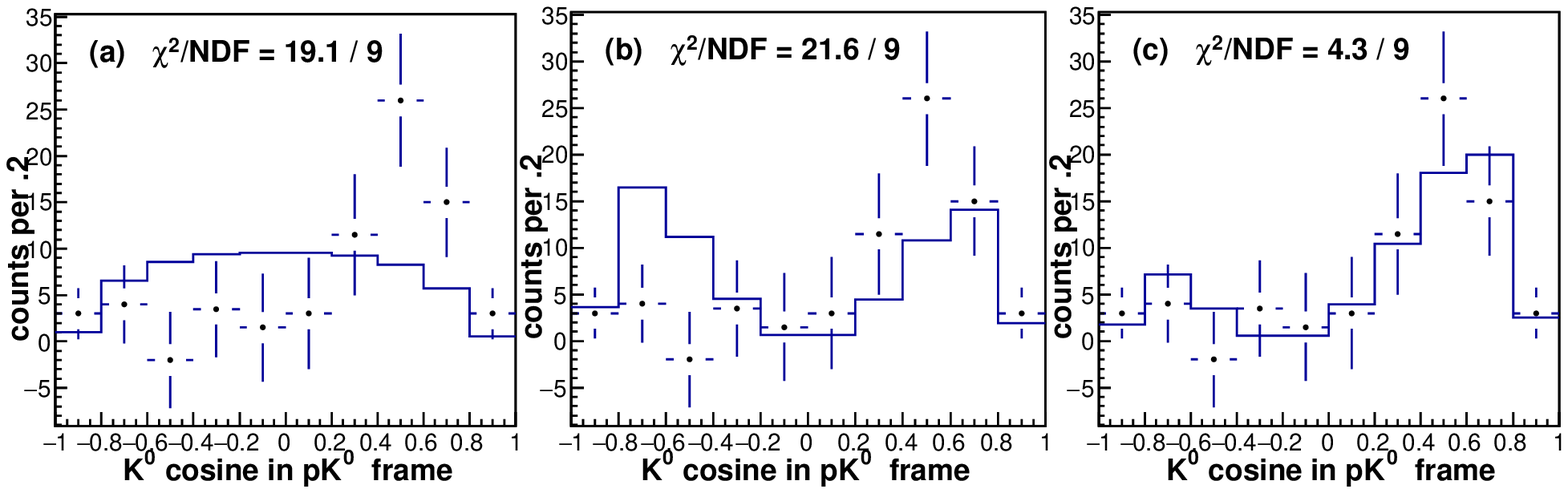}
\caption {
Under the selections \alotra, the sideband-subtracted \cosdec\ distribution 
for the $\Theta^+$ mass region compared with Monte-Carlo predictions for the 
$\Theta^+$ decay angular distribution in the assumed forms
$dW/d\cos\Theta_K^\mathrm{cms} \sim \mathrm{const.}$ (a), 
\symme\ (b), and \askew\ with $a = +0.2$ and $b = 0$ (c). The simulated distributions 
have been scaled to the data by area.}
\label{pure}
\end{figure}

     In order to obtain the ``pure" \cosdec\ spectrum for the decay 
$\Theta^+ \rightarrow pK^0$, we subtract the (halved) \cosdec\ 
distribution for the sidebands from that for the $\Theta^+$ peak region.
Under the selections \alotra\ that reject nearly a half of the rescattered 
events, the sideband-subtracted \cosdec\ spectrum is shown in Fig.~\ref{pure}.
It is then compared with Monte-Carlo predictions for the $\Theta^+$ formation
and decay assuming the angular distribution in the forms 
$dW/d\cos\Theta_K^\mathrm{cms} \sim \mathrm{const.}$ and \symme\ as shown in 
Figs. \ref{pure}(a) and \ref{pure}(b), where the simulated distributions have 
been scaled to the data by area. These forms are the two allowed components
of the \cosdec\ spectrum for arbitrary spin--parity of the decaying $\Theta^+$ 
baryon. For both hypotheses, the values of $\chi^2/\mathrm{ndf}$ (19.1/9 and 
21.6/9) are unacceptably high. We have also verified numerically that any 
linear combination of the above forms of the \cosdec\ distribution for the
decay $\Theta^+ \rightarrow pK^0$ leads to $\chi^2$ values in excess of 
19 when compared with the observed sideband-subtracted \cosdec\ spectrum.

     In order to verify the conclusions reached with the $\chi^2$ analysis
of our low-statistics data, we also use the likelihood criterion.
The likelihood function is constructed as a sum
\begin{displaymath}
-2\ln L = 2 \sum_{i=1}^{10} [ -n_i + \nu_i + h_i + n_i \ln \frac{n_i}{\nu_i + h_i}] ,
\end{displaymath}
where $n_i$ and $\nu_i$ are the bin contents of the experimental and simulated
\cosdec\ distributions for the $\Theta^+$ central mass region shown in 
Fig.~\ref{peak-wings}(b),  and $h_i$ are those of the simulated \cosdec\ 
distribution for the given hypothesis of $\Theta^+$ decay. The latter
distribution has been scaled by the number of events to the difference between 
the former two, so that the fitting function as a whole is normalized
to the observed \cosdec\ distribution by area. For the hypotheses 
$dW/d\cos\Theta_K^\mathrm{cms} \sim \mathrm{const.}$ and \symme, 
we obtain $-2\ln L = 24.2$ and 23.1, respectively. These values of $-2\ln L$ for 
$\mathrm{ndf} = 9$ correspond to the $p$-values near 0.005 and .007. As with the 
$\chi^2$ analysis above, we find that assuming the $\Theta^+$ decay angular 
distribution in the form of an arbitrary linear combination of the former two
fails to tangibly reduce the value of $-2\ln L$. This simple analysis
supports the above conclusions based on the $\chi^2$ criterion. From the derived
$p$-values we may conclude that the uniform angular distribution
$dW/d\cos\Theta_K^\mathrm{cms} \sim \mathrm{const.}$ is inconsistent with the 
data at a statistical level of 2.9$\sigma$, and the more general symmetric form
$dW/d\cos\Theta_K^\mathrm{cms} \sim a + b\cos^2\Theta_K^\mathrm{cms}$ --- at a
slightly lower level of 2.8$\sigma$.
\begin{figure}[!b]
\renewcommand{\baselinestretch}{1.}
\vspace { 15.2 cm }
\includegraphics{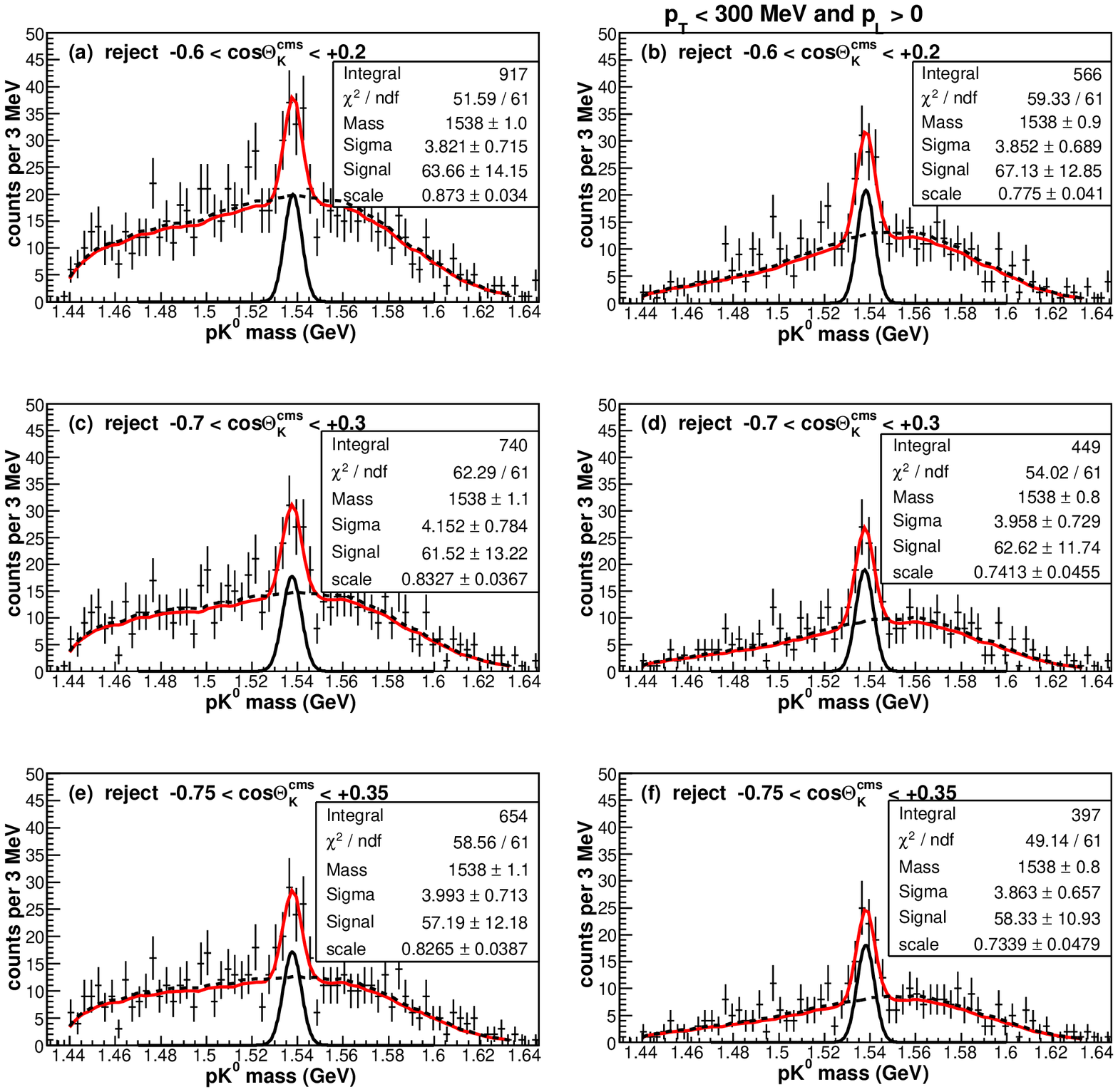}
\caption {
The $pK^0$ effective-mass spectrum upon rejecting the events 
that fall within the \cosdec\ intervals of \one\ (a), \two\ (c), and 
\three\ (e). The effect of the selections \alotra\ is shown in the 
right-hand panels (b), (d), and (f). Each mass spectrum is fitted to the 
simulated nonresonant background with variable normalization plus a 
Gaussian with variable position, width, and magnitude. The null fits to 
the background form alone are shown by dashed lines.}
\label {cut-cosdec}
\end{figure}

     The disagreement with any viable form of the $\Theta^+$ decay angular
distribution is rooted in the marked forward-backward asymmetry of the
observed \cosdec\ spectrum. The asymmetry in the form of a linear term
in \cosdec, required at a level of 2.8$\sigma$, may arise only from the 
interference \cite{Sibirtsev} between
the nonresonant $s$-wave and the $\Theta^+$-mediated higher-wave contributions
to the amplitude of the charge-exchange reaction $K^+n \rightarrow pK^0$.
However, the interference should not affect the \cosdec\ distribution for the
signal as soon as the $\Theta^+$ spin-parity is 1/2$^-$ implying an $s$-wave
decay and a uniform \cosdec\ distribution. Therefore, the 1/2$^-$ assignment 
is ruled out by the data at a level of 2.9 standard deviations. 

     The shape of the background-subtracted $\Theta^+$ signal in Fig. \ref{pure}
suggests an angular dependence of an asymmetric quadratic form \askew, where 
$b \simeq 0$ and the offset parameter $a$ is positive at a 2.8$\sigma$ level. 
Tentatively substituting $b = 0$ and
$a = +0.2$, we obtain $\chi^2/\mathrm{ndf} = 4.3/9$ with the $\chi^2$ approach
as shown in Fig.~\ref{pure}(c), and $-2\ln L = 6.1$ for $\mathrm{ndf} = 9$ with
the likelihood criterion which corresponds to a $p$-value near 0.8. (The 
assignments $a = +0.1$ and $a = +0.3$ also yield acceptable values of $\chi^2$
and $-2\ln L$.) The detailed interpretation of this form of the 
background-subtracted \cosdec\ distribution requires a theoretical analysis of 
interference effects in terms of helicity amplitudes, and therefore is beyond 
the scope of this paper (apart from excluding the $\Theta^+$ spin--parity 
assignment of 1/2$^-$). Instead, our major objective is to formulate a 
physically-reasonable data-driven
selection that may render the $\Theta^+$ signal more significant. This is an
important task since the signals reported thus far \cite{Nakano, DIANA-2014,
Amaryan} are but slightly in excess of 5$\sigma$, whereas the ``credibility 
threshold" for proving the $\Theta^+$ existence has been estimated as 
7$\sigma$ \cite{Lions} given the controversial experimental situation.

     If formation of the $\Theta^+$ baryon indeed follows the 
distribution \askew\ as argued above, rejecting the events with 
\cosdec\ values near the minimum of the parabola at $-a$ should enhance 
the signal-to-background ratio in the $pK^0$ effective-mass spectrum and
the statistical significance of the $\Theta^+$ peak.
Shown in Fig.~\ref{cut-cosdec} are the effects of cutting away the
\cosdec\ intervals centered on \cosdec\ = -0.2 : \one, \two, and \three. 
Despite the uncertainty of the angular-distribution parameters, these 
selections are physically meaningful rather than arbitrary.
Each mass spectrum is again fitted to the simulated nonresonant background 
with variable normalization plus a Gaussian with variable position, width, 
and magnitude. The width of the observed $\Theta^+$ peak is always 
consistent with the experimental resolution of $\sigma_m \simeq 3.5$ MeV. 
Indeed, cutting on \cosdec\ is seen to result in a dramatic increase of the 
signal-to-background ratio as compared to the $pK^0$ mass spectra 
of Fig.~\ref{dimass}.
\begin{table}[t]
\small
\begin{tabular}{|l|l|l|l|l|c|c|}
\hline
Rejected \cosdec\ interval & $m_0$ (MeV) & Signal (ev) & $-\ln L$     & $-\ln L$ & $2\Delta\ln L$   & Stat. \\
                           &             &$S/\sqrt{B}$ & $\chi^2/$ndf & $\chi^2/$ndf &              & sign. \\
                           &             &             & (signal fit) & (null fit)   &         &            \\
\hline
\hline
None                       & $1538\pm1$  & $74.9\pm14.5$& 31.5        & 46.6         & 30.1    &  5.1$\sigma$ \\
                           &             &  6.8         & 64.5/62     & 91.3/64      &         &        \\
\hline
   \one\                   & $1538\pm1$  & $64.1\pm11.3$& 35.0        & 57.2         & 44.4    &  6.3$\sigma$ \\
                           &             &  8.6         & 59.6/62     & 91.9/64      & &                \\
\hline
     \two\                 & $1538\pm1$  & $59.3\pm10.3$& 30.9        & 55.2         & 48.5 &     6.6$\sigma$ \\
                           &             & 9.3          & 54.4/62     & 87.5/64      &      &           \\
\hline
     \three\               & $1538\pm1$  & $55.8\pm9.8$ & 29.3        & 54.4         & 50.2 &     6.8$\sigma$ \\
                           &             & 9.4          & 49.5/62     & 81.8/64      &      &            \\
\hline
\end{tabular}
\renewcommand{\baselinestretch}{1.}
\caption
{The results of the fits of the $pK^0$ mass spectra under the 
selections \alotra, in which the Gaussian width of the signal has been 
constrained to the simulated resolution of $\sigma_m = 3.5$ MeV. 
The statistical significance of the signal, estimated using the method
of maximum likelihood, is shown in the rightmost column. Also
shown is the ``naive" estimate of the statistical significance $S/\sqrt{B}$, 
where the signal $S$ and the background $B$ are derived from the signal 
hypothesis alone over the 90\% area of the Gaussian.}
\label{constrained}
\end{table}

     In order to reduce the number of free parameters, the width of the
peak is constrained to the simulated value of $\sigma_m = 3.5$ MeV when 
estimating the statistical significance of the signal. The results of the 
constrained fits of the $pK^0$ mass spectra under
the selections \alotra\ are shown in Table \ref{constrained}. Also shown 
for each fit is the difference between the log-likelihood values for the 
signal and null hypotheses, $-2\Delta\ln L$. For the constrained fits,
the numbers of degrees of freedom for the signal and null hypotheses differ
by $\Delta\mathrm{ndf} = 2$. The statistical significance of the signal is 
estimated using the value of $\chi^2$ for one degree of freedom which 
corresponds to the same $p$-value as $\chi^2 = -2\Delta\ln L$ for two
degrees of freedom. Rejecting the central values of \cosdec\ is seen to 
boost the statistical significance of the $\Theta^+$ signal from 
5.1$\sigma$ up to 6.8$\sigma$. The ``naive" estimate of the statistical
significance reaches $S/\sqrt{B} = 9.4\sigma$, where the signal $S$ and the
background $B$ have been derived from the signal fit alone over the 90\%
area of the Gaussian. That the significance of the $\Theta^+$ signal is 
substantially increased by an asymmetric \cosdec\ cut {\it a posteriori} 
indicates that both the quadratic and linear terms contribute to the \cosdec\
distribution.

\begin{figure}[t]
\renewcommand{\baselinestretch}{1.}
\vspace { 11 cm }
\includegraphics{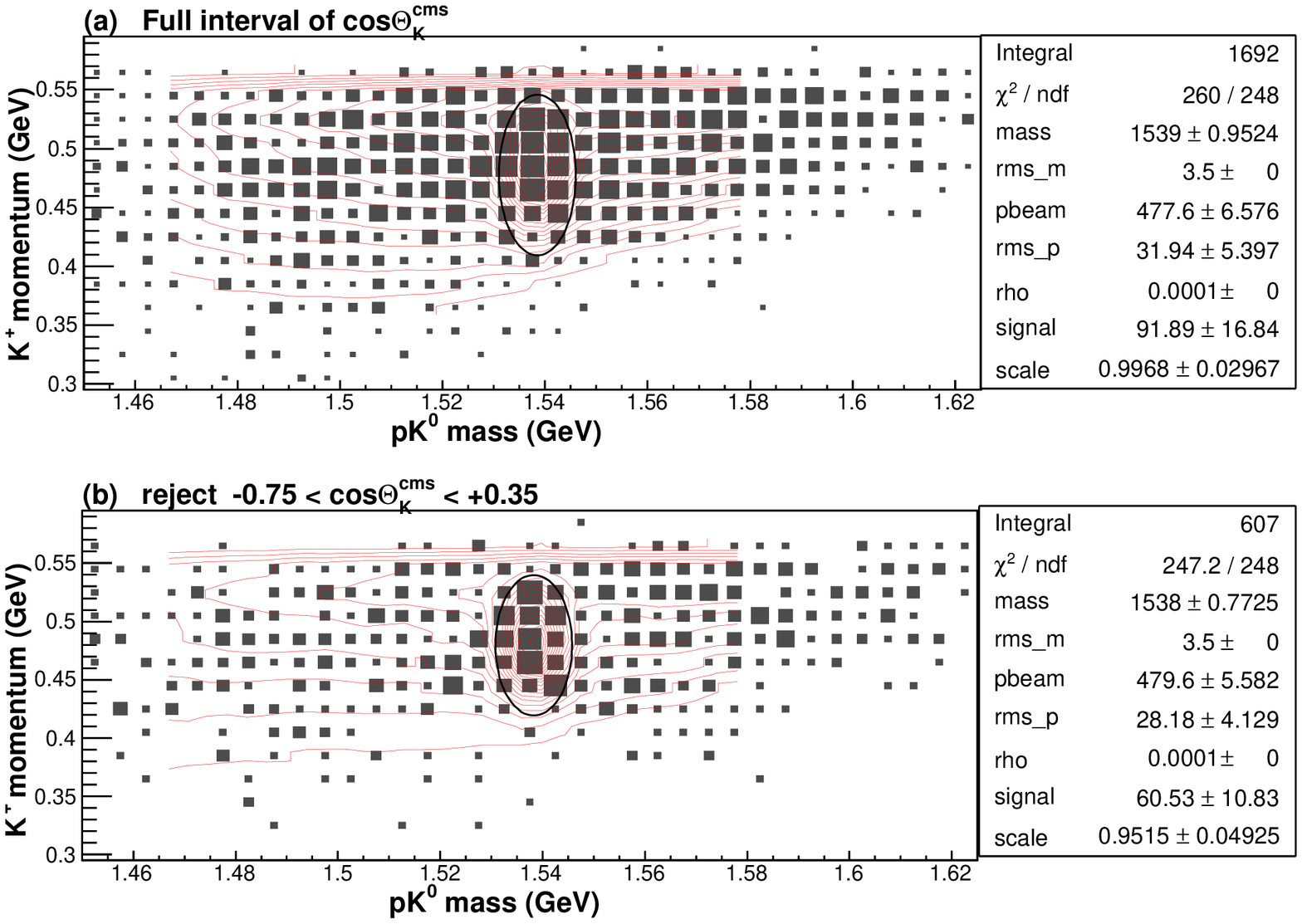}
\caption {
Under the selections \alotra, beam momentum $p(K^+)$ 
plotted versus the $pK^0$ effective mass (a). The effect of rejecting the 
events with \three\ is shown in (b). Either scatter plot is fitted to 
the corresponding simulated distribution with variable normalization plus a 
two-dimensional Gaussian. The width of the Gaussian in \dimass\ and the 
correlation parameter $\rho$ have been constrained to the experimental mass 
resolution $\sigma_m = 3.5$ MeV and to zero, respectively. The ellipses show
the 90\% areas of the Gaussians.}
\label{scaplo}
\end{figure}
     And finally, beam momentum $p(K^+)$ is plotted versus the $pK^0$ effective 
mass inder the selections \alotra\ in Fig.~\ref{scaplo}(a), and upon rejecting
the events with \three\ --- in Fig.~\ref{scaplo}(b). The latter selection results 
in a distinct $\Theta^+$ signal at expected values of \dimass\ and $p(K^+)$. Either
scatter plot is then fitted to the corresponding simulated distribution with 
variable normalization plus a two-dimensional Gaussian. The width of the Gaussian
in \dimass\ is constrained to the experimental mass resolution of $\sigma_m = 3.5$
MeV, and the correlation parameter $\rho$ is constrained to zero as physically
expected for formation of a narrow resonance (as the observed mass should not depend 
on beam momentum). Either scatter plot has also been fitted to the background form
alone (not shown). For the fits in Figs. \ref{scaplo}(a) and \ref{scaplo}(b), we
have $-2\Delta\ln L = 49.7$ and 61.9 for $\Delta\mathrm{ndf} = 4$. The 
statistical significance of the signal is again estimated using the value of 
$\chi^2$ for one degree of freedom which corresponds to the same $p$-value as 
$\chi^2 = -2\Delta\ln L$ for four degrees of freedom. Thereby, we obtain that
cutting away the \cosdec\ region of \three\ allows to increase the statistical 
significance of the $\Theta^+ \rightarrow pK^0$ signal from 6.2$\sigma$ up to 
7.1$\sigma$.

     In summary, using the data on the charge-exchange reaction 
$K^+n \rightarrow pK^0$ on a bound neutron, we have analyzed the dependence of the 
background-subtracted $\Theta^+ \rightarrow pK^0$ signal on the $K^0$ emission 
angle in the $pK^0$ rest frame, $\Theta_K^\mathrm{cms}$. In order to describe the 
observed \cosdec\ distribution, invoking the interference between the nonresonant 
$s$-wave and the $\Theta^+$-mediated higher-wave contributions to the amplitude 
of the charge-exchange reaction is required at a level of 2.8$\sigma$. The 
spin--parity assignment of 1/2$^-$ for the $\Theta^+$ baryon is ruled out at a 
statistical level of 2.9$\sigma$. A physically-meaningful selection in \cosdec\ 
based on the observed angular dependence 
of the background-subtracted $\Theta^+ \rightarrow pK^0$ signal allows to 
boost the statistical significance of the signal up to 6.8$\sigma$ for the 
one-dimensional $pK^0$ mass spectrum, and 7.1$\sigma$ for the scatter plot in 
\dimass\ and $p(K^+)$. This is far in excess of previously reported signals 
\cite{Nakano, DIANA-2014, Amaryan} and renders the $\Theta^+$ existence 
more credible \cite{Lions}. A high-statistics investigation of the charge-exchange
reaction $K^+n \rightarrow pK^0$ is needed for finally proving the $\Theta^+$
existence and fixing its quantum numbers.

     Instructive discussions with professor Ya. Azimov of the St. Petersburg 
Institute of Nuclear Physics are gratefully acknowledged.

\end{document}